\documentclass[12pt]{iopart}
\usepackage{epsfig}
\usepackage{hyperref}

\begin{document}

\begin{flushright}
WITS-MITP-023
\end{flushright}

\title[]{The evolution of gauge couplings and the Weinberg angle in 5 dimensions for an SU(3) gauge group}

\author{{Mohammed Omer khojali}$^{a,1}$, {A. S. Cornell}$^{a,2}$ and 
{Aldo Deandrea}$^{b,3}$ }

\address{ $^{a}$ National Institute for Theoretical Physics; School of Physics and Mandelstam Institute for
Theoretical Physics, University of the Witwatersrand, Wits 2050, South Africa\\
$^{b}$ Universite de Lyon, F-69622 Lyon, France; Universite Lyon 1, CNRS/IN2P3, UMR5822 IPNL, F-69622 Villeurbanne Cedex, France}
\ead{ $^{1}$khogali11@gmail.com, $^{2}$alan.cornell@wits.ac.za, $^{3}$deandrea@ipnl.in2p3.fr}
\vspace{10pt}
\begin{indented}
\item[]
\end{indented}

\begin{abstract}
 We test in a simplified 5-dimensional model with SU(3) gauge symmetry, the evolution equations of the gauge couplings of a 
 model containing 
bulk fields, gauge fields and one pair of fermions. In this model we assume that the fermion doublet and two singlet 
fields are located at fixed points of the extra-dimension compactified on an $S^{1}/Z_{2}$ orbifold.
The gauge coupling evolution is derived at one-loop in 5-dimensions, for the gauge group
$G = SU(3)$, and used to test the impact on lower energy observables, in particular the
Weinberg angle. 
The gauge bosons and the Higgs field arise from the gauge bosons in 5 dimensions, as in a gauge-Higgs model.
The model is used as a testing ground as it is not 
a complete and realistic model for the electroweak interactions.
\end{abstract}

%
%
%
%
%

\section{Introduction}
A gauge theory defined in more than four dimensions has many attractive features, where interactions at
low energies may be truely unified and some of the distinct fields in four dimensions can be integrated as a single multiplet in 
higher dimensions, like in gauge-Higgs models, where the Higgs fields 
can be a component of 5-dimensional gauge fields. Note also that the  topology and structure of the
extra-dimensional space provides new ways of breaking symmetries~\cite{R0}. The simplest theories of this type
have problems in reproducing the low energy observables, such as the Weinberg angle, the SM fermion content and 
Yukawa couplings are different from the gauge couplings~\cite{R1}.\par
In this paper we shall discuss the gauge couplings evolution for a model which contains a bulk field, gauge fields and one pair of
fermions $\psi_{a}$ and $\tilde{\psi}_{a}$. The matter field can be introduced either as a bulk field in the representations
 of the unified group $G = SU(3)$ or as a boundary field localised at the fixed point where this is broken to a subgroup $H$.\par
 Let us call $H$ the subgroup of $G$ that is not broken by the vacuum expectation value (vev) of the scalar fields 
 (under which the vev of the scalar fields is invariant). We can correspondingly divide the generators of $G$ into two sets:
the unbroken ones $SU(2) \times U(1)$ (the electroweak gauge group), which annihilates the vacuum, and the broken ones $\hat{U}(1)$ 
(the electromagnetic group), the orthogonal set. According to the Goldstone theorem each broken generator in the coset $G/H$ is
associated to an independent massless scalar (Goldstone bosons), carrying the same quantum numbers as the generators~\cite{R2}:
\begin{equation}
G/H = \frac{SU(3)}{SU(2)\times U(1)} \Rightarrow dim ($G/H$) = 8 - (3+1)= 4.
\end{equation}\par
In the case of the bulk fields, the standard Yukawa coupling can orginate only from higher-dimensional gauge couplings, but in 
the case of the boundary localised matter fields the standard Yukawa coupling cannot be directly introduced~\cite{R3}. 
The gauge bosons arise from the 4-dimensional components of the 5-dimensional gauge fields, whilst the Higgs field
arises from the internal components of the gauge group $G = SU(3)$ compactified on an $S^{1}/Z_{2}$ orbifold;
the orbifold boundary condition can be written in the following way
\begin{equation}
P = e^{i\pi \lambda_{3}} =
\left( \begin{array}{cccc}
-1&0&0 \\
0&-1&0 \\
0&0&1\\
\end{array} \right),
\end{equation}
where $\lambda_{a}$ are the standard $SU(3)$ Gell-Mann matrices, normalised as $Tr (\lambda_{a}\lambda_{b}) = 2\delta_{ab}$.
 The group $Z_{2}$ acts on the tours as $\pi$ rotations, the orbifold projection $P$ breaking the gauge
 group $G$ to the subgroup $H$ = $SU(2)\times U(1)$, 
the group $G$ is broken in 4 dimensions to $H$ = $SU(2)\times U(1)$ of the projection $P$, the massless 4-dimensional fields are 
the gauge bosons $A_{\mu}^{a}$ in the adjoint of H and the charged scalar doublet arises from the internal components
$A_{5}^{a}$ of the gauge field~\cite{R4}.\par
 The brane fields of the model we shall focus on consist of a left-handed fermion doublet $Q_{L} = ( u_{L},d_{L})$, and
 two right-handed fermion singlets $u_{R}$ and $d_{R}$.  We are going to assume that the doublet and the two 
 singlet fields are located respectively at position $y_{1}$ and $y_{2}$, which equals to either 0 or $\pi$R.\par
The Lagrangian for the bulk fields, gauge fields and the pair of fermions is given by: 
\begin{eqnarray}
 \mathcal{L}_{matter} =& \sum_{a}\Bigg[ i\bar{\psi}_{a}(x,y) \displaystyle{\not}D_{5} \psi_{a}(x,y)
 +i\bar{\tilde{\psi}}_{a}(x,y) 
 \displaystyle{\not}D_{5} \tilde{\psi}_{a}(x,y)
  \nonumber\\& + \bar{\psi}_{a}(x,y)M_{a}\tilde{\psi}_{a}(x,y) + \bar{\tilde{\psi}}_{a}(x,y)M_{a}\psi_{a}(x,y)\Bigg]
 \nonumber\\&  + 
 \delta(y -y_{1})\Bigg[i\tilde{Q}_{L}(x,y)\displaystyle{\not}D_{\mu}Q_{L}(x,y)\Bigg]\nonumber\\&+
 \delta(y-y_{2})\Bigg[i \bar{d}_{R}(x,y)\displaystyle{\not}D_{\mu}d_{R}(x,y)
 + i\bar{u}_{R}(x,y)\displaystyle{\not}D_{\mu}u_{R}(x,y)\Bigg],
\end{eqnarray}
where $\displaystyle{\not}D_{4}$ and 
$\displaystyle{\not}D_{5}$ are the 4-dimensional and 5-dimensional covariant derivatives respectively, and are related by the 
following equality 
\begin{equation}
 \displaystyle{\not}D_{5} = \displaystyle{\not}D_{4} + i \gamma_{5}D_{5}.
 \end{equation}
 \begin{equation}
  \displaystyle{\not}D_{M} = \gamma^{M}\partial_{M} - i\gamma^{M} g_{M}A^{a}_{M}T^{a},
 \end{equation}
where M = ($\mu\equiv 0,1,2,3 $ and 5), the Hermitian martix $\gamma_{5} = i\gamma_{4}$,
 $T^{a}$ are the generators of the Lie algebra of the gauge group $G$, $A_{\mu}^{a}$ 
 are the 4-dimensional gauge bosons and the scalar fields $A_{5}^{a}$ are identified with the components of the Higgs field H
 ~\cite{R5}.\par
 In the fundamental representation of the gauge group $G$, the mode expansion
 for the left-handed $\psi_{L}$ and the right-handed $\psi_{R}$ bulk fermion is 
\begin{equation}~\label{eq:M1}
  \psi_{aL}(y) = \sum_{-\infty}^{\infty} \eta_{n}\frac{1}{\sqrt{2\pi R}} \sin\Bigg(\frac{ny}{R}\Bigg) \psi^{n}_{aL}(x),
 \end{equation}
 \begin{equation}~\label{eq:M2}
  \psi_{aR}(y) = \sum_{-\infty}^{\infty} \eta_{n}\frac{1}{\sqrt{2\pi R}} \cos\Bigg(\frac{ny}{R}\Bigg) \psi^{n}_{aR}(x).
 \end{equation}
By adding equations ~(\ref{eq:M1}) and ~(\ref{eq:M2}) one can get the corresponding Fourier decomposition of
a generic bulk fermion
\begin{equation}~\label{eq:M3}
 \psi_{a}(y) = \sum_{n= -\infty}^{\infty} \eta_{n}\frac{1}{\sqrt{2\pi R}} \Bigg[\sin\Bigg(\frac{ny}{R}\Bigg)\psi^{n}_{aL}(x)+
 \cos\Bigg(\frac{ny}{R}\Bigg)\psi^{n}_{aR}(x)\Bigg],
\end{equation}
where the factor $\eta_{n}$ is defined to be 1 for n = 0 and 1/$\sqrt{2}$ for $n \neq 0 $, which means we can rewrite the bulk 
fermion in equation~(\ref{eq:M3}) as
\begin{equation}
 \psi_{a}(y) = \frac{1}{\sqrt{2\pi R}} \psi^{0}_{aR}(x) + \frac{1}{\sqrt{2\pi R}}\sum_{n = 1}^{\infty}
 \Bigg[\sin\Bigg(\frac{ny}{R}\Bigg) \psi^{n}_{aL}(x) + \cos\Bigg(\frac{ny}{R}\Bigg)\psi^{n}_{aR}(x)\Bigg].
\end{equation}\par
The 4-dimensional Lagrangian for the bulk fermion $\psi_{a}$ is written as
\begin{equation}
 \mathcal{L}_{4D}^{\psi_{a}} = \int_{0}^{\pi R} \Bigg[\bar{\psi}_{a}(y)i \displaystyle{\not}D_{5} \psi_{a}(y)\Bigg] dy ,
\end{equation}
where integrating out the $y$ coordinate one can get
\begin{eqnarray}~\label{eq:M4}
 \mathcal{L}_{4D}^{\psi_{a}} =& \frac{1}{2}\bar{\psi}^{0}_{aR}(x)(\gamma^{4}\partial_{4} - \gamma_{4}\partial_{5})
 \psi_{aR}^{0}(x) + \frac{i\gamma_{4}g_{5}T^{a}}{2} \bar{\psi}_{aR}^{0}(x) A_{5}^{a} \psi_{aR}^{0}(x)\nonumber \\& 
 - \frac{i\gamma^{4}g_{4}T^{a}}{2}\bar{\psi}_{aR}^{0}(x) A_{4}^{a} \psi_{aR}^{0}(x) + 
 \frac{1}{8}\sum_{n=1}^{\infty}\bar{\psi}^{n}_{aL}(x)(\gamma^{4}\partial_{4} - \gamma_{4}\partial_{5}) \psi^{n}_{aL}(x)
 \nonumber\\& + \frac{i\gamma_{4}g_{5}T^{a}}{8}\sum_{n=1}^{\infty}\bar{\psi}^{n}_{aL}(x)A^{a}_{5}\psi^{n}_{aL}(x) -
 \frac{i\gamma^{4}g_{4}T^{a}}{8}\sum_{n=1}^{\infty} \bar{\psi}^{n}_{aL}(x)A^{a}_{4}\psi^{n}_{aL}(x)\nonumber\\&
 +\frac{1}{8}\sum_{n=1}^{\infty} \bar{\psi}^{n}_{aR}(x)(\gamma^{4}\partial_{4} - \gamma_{4}\partial_{5})\psi^{n}_{aR}(x)
 + \frac{i\gamma_{4}g_{5}T^{a}}{8}\sum_{n=1}^{\infty}\bar{\psi}^{n}_{aR}(x)A^{a}_{5}\psi^{n}_{aR}(x)\nonumber\\&
 - \frac{i\gamma^{4}g_{4}T^{a}}{8}\sum_{n=1}^{\infty} \bar{\psi}^{n}_{aR}(x)A^{a}_{4}\psi^{n}_{aR}(x).
\end{eqnarray}
We can obtain the 4-dimensional Lagrangian for the bulk fermion $\tilde{\psi}_{a}$, in 
 similar way as in the case of the bulk fermion $\psi_{a}$, by replacing $\psi_{a}$ by $\tilde{\psi}_{a}$
 in equation ~(\ref{eq:M4}).\par
 Now let us move to the case of the 4-dimensional left-handed fermion doublet, where the Fourier decomposition for that field  is 
 written as
\begin{equation}
  Q_{L}(y) = \frac{1}{\sqrt{2\pi R}}Q^{0}_{L}(x) + \frac{1}{2\sqrt{\pi R}}\sum_{n=1}^{\infty}\Bigg[\cos\Bigg(\frac{ny}{R}\Bigg)
  Q_{L}^{n}(x) + \sin\Bigg(\frac{ny}{R}\Bigg)Q^{n}_{R}(x)\Bigg].
 \end{equation}\par
 The 4-dimensional Lagrangian for the left-handed fermion doublet is given by
\begin{equation}
  \mathcal{L}^{Q_{L}}_{4D} = \int_{0}^{\pi R} dy \delta(y-y_{1})\Bigg[\bar{Q}_{L}i\displaystyle{\not}D_{4}Q_{L}\Bigg],
 \end{equation}
where as we mentioned before, the $\delta(y - y_{1})$ is needed as the left-handed fermion doublet is located at position $y_{1}$, 
which is equal
to either 0 or $\pi R$. By integrating out the $y$ coordinate one can get
\begin{eqnarray}
  \mathcal{L}^{Q_{L}}_{4D} =& \frac{1}{4\pi R}\Bigg[\bar{Q}_{L}^{0}(x)[ i\gamma^{4}\partial_{4} + \gamma^{4}g_{4}A_{4}^{a}T^{a}] 
  Q_{L}^{0}(x)\nonumber\\&  + \frac{1}{2} \sum_{n=1}^{\infty} \bar{Q}^{n}_{L}(x)[i\gamma^{4}\partial_{4} + 
  \gamma^{4}g_{4}A_{4}^{a}T^{a}] Q^{n}_{L}(x)\Bigg].
 \end{eqnarray}
\par Finally, we can see the case of the two singlet fields which are located at position $y_{2}$,  the Fourier
decomposition for those fields are written as
 \begin{equation}
  d_{R}(y) = \frac{1}{\sqrt{2\pi R}} d^{0}_{R}(x) + \frac{1}{2\sqrt{\pi R}}\sum_{n=1}^{\infty}
  \Bigg[\cos\Bigg(\frac{ny}{R}\Bigg)d^{n}_{R}(x) + \sin\Bigg(\frac{ny}{R}\Bigg)d^{n}_{L}(x)\Bigg],
 \end{equation}
\begin{equation}
  u_{R}(y) = \frac{1}{\sqrt{2\pi R}} u^{0}_{R}(x) + \frac{1}{2\sqrt{\pi R}}\sum_{n=1}^{\infty}
  \Bigg[\cos\Bigg(\frac{ny}{R}\Bigg)u^{n}_{R}(x) + \sin\Bigg(\frac{ny}{R}\Bigg)u^{n}_{L}(x)\Bigg].
 \end{equation}
\par The 4-dimensional Lagrangian for the two singlet fields $d_{R}$ and $u_{R}$ is written as
 \begin{equation}
  \mathcal{L}_{4D}^{singlet} = \int_{0}^{\pi R} dy \delta(y-y_{2})
  \Bigg[\bar{d}_{R}i\displaystyle{\not}D_{4}d_{R} + \bar{u}_{R}i\displaystyle{\not}D_{4}u_{R}\Bigg],
 \end{equation}

 where by integrating out the $y$ coordinate one can get 
 \begin{eqnarray}
  \mathcal{L}_{4D}^{singlet}=& \frac{1}{4\pi R}\Bigg[ \bar{d}_{R}^{0}(x)[i\gamma^{4}\partial_{4} + 
  \gamma^{4}g_{4}A^{a}_{4}T^{a}]d^{0}_{R}(x)\nonumber\\& + \bar{u}^{0}_{R}(x)
  [i\gamma^{4}\partial_{4} + \gamma^{4}g_{4}A^{a}_{4}T^{a}]u^{0}_{R}(x)\nonumber\\&
  +\frac{1}{2}\sum_{n=1}^{\infty}\bar{d}^{n}_{R}(x)[i\gamma^{4}\partial_{4} + \gamma^{4}g_{4}A^{a}_{4}T^{a}]
  d^{n}_{R}(x)\nonumber\\& + \frac{1}{2}\sum_{n=1}^{\infty}\bar{u}^{n}_{R}(x)[i\gamma^{4}\partial_{4} +
  \gamma^{4}g_{4}A^{a}_{4}T^{a}]u^{n}_{R}(x)\Bigg].
 \end{eqnarray}
\section{The gauge coupling evolution equations}
Our goal is to discuss the gauge coupling evolution for the model presented in the previous section. 
 In order to do so, we need to introduced the $\beta$-functions.
This crucial object is needed to determine the evolution of the coupling constants. In general, in a theory with $n$-couplings 
$g_{i}$, we have to solve a set of coupled differential equations of the form
\begin{equation}
 \beta_{i} = \mu\frac{d g_{i}}{d \mu} = \frac{d g_{i}}{d t},
\end{equation}
where $t = (\ln[\mu/M_{Z}])$. In general the $\beta$-functions depend on all the couplings and masses of the theory.
We can get rid of the masses by focusing only on the universal UV relevant coefficients. For example, one can focus on 
the gauge coupling evolution equations, where we can write the general term for the gauge interaction of the fermions and the gauge
bosons as $ g\bar{\psi}\gamma^{\mu}\psi A_{\mu}$.
 In terms of renormalisable quantities (by rescaling)
\begin{equation}~\label{eq:2}
 \bar{\psi} = Z^{1/2}_{\bar{\psi}} \bar{\psi}^{R},
\end{equation}
\begin{equation}~\label{eq:3}
 \psi = Z_{\psi}^{1/2} \psi^{R},
\end{equation}
\begin{equation}~\label{eq:4}
 A_{\mu} = Z_{A_{\mu}}^{1/2}A_{\mu}^{R},
\end{equation}
where $Z_{\psi}^{1/2}$, $Z^{1/2}_{\bar{\psi}}$ and $Z_{A_{\mu}}^{1/2}$ are the renormalisation constants. By using equations
~(\ref{eq:2}), ~(\ref{eq:3}) and ~(\ref{eq:4}) one can write the gauge interaction of the fermions and the gauge bosons 
 in terms of the renormalisable quantities
\begin{equation}
 gZ^{1/2}_{\bar{\psi}}Z^{1/2}_{\psi}Z^{1/2}_{A_{\mu}} \bar{\psi}^{R}\gamma^{\mu}\psi^{R}A_{\mu}^{R} = Z^{1/2}_{g}g^{R}
 \bar{\psi}^{R}\gamma^{\mu}\psi^{R}A_{\mu}^{R}.
\end{equation}
From the above equation one can see that 
\begin{equation}
 gZ^{1/2}_{\bar{\psi}}Z^{1/2}_{\psi}Z^{1/2}_{A_{\mu}} = Z^{1/2}_{g}g^{R}.
\end{equation}
As we discussed earlier, the couplings $g_{i}$ are determined by noticing that physics cannot depend on our 
arbitrary choice of scale $\mu$. We have, therefore,
\begin{equation}
 \frac{d \ln g^{R}}{d t} = \frac{1}{2}\frac{d \ln Z_{\bar{\psi}_{R}}}{d t} + 
 \frac{1}{2}\frac{d \ln Z_{\psi_{R}}}{d t} + \frac{1}{2}\frac{d \ln Z_{A_{\mu}}}{d t } -
 \frac{1}{2}\frac{d \ln Z_{g}}{d t}.
\end{equation}
We then need to calculate the renormalisation constants. When doing so, we usually ignore
the mass terms in the propagators, since they have nothing to do with the divergent part of the one loop diagrams.
 We are going to focus on the UV regime where we can neglect the
$m/\mu$ dependence of $\beta$.\par 
The general formula of the $\beta$-functions for the gauge couplings is given by:
\begin{equation}~\label{R22}
 16\pi^{2} \frac{d g_{i}}{dt} = b_{i}^{SM} g_{i}^{3} +( b_{i} + S(t) \tilde{b_{i}})g_{i}^{3},
\end{equation}
where $t = \ln(S(t)/M_{Z}R)$, $S(t) = \mu R$ for $M_{Z} < \mu < \ln(1/M_{Z}R)$.
The numerical coefficients appearing in equation~(\ref{R22}) are given by:
\begin{eqnarray}
 b_{i}^{SM} = \Bigg[ \frac{41}{10}, -\frac{19}{6}, -7 \Bigg],\quad 
 b_{i} = \Bigg[\frac{10}{3}, -\frac{51}{16}, -\frac{20}{3}, \frac{3}{8} \Bigg], \quad
  \tilde{b_{i}} = \Bigg[ \frac{45}{16\pi}, -\frac{15}{16\pi}, -\frac{5}{\pi}, 0 \Bigg]\nonumber \\.
\end{eqnarray}
\section{Result and discussion}
\begin{figure}[h]
\centering
\includegraphics[width=6.5cm]{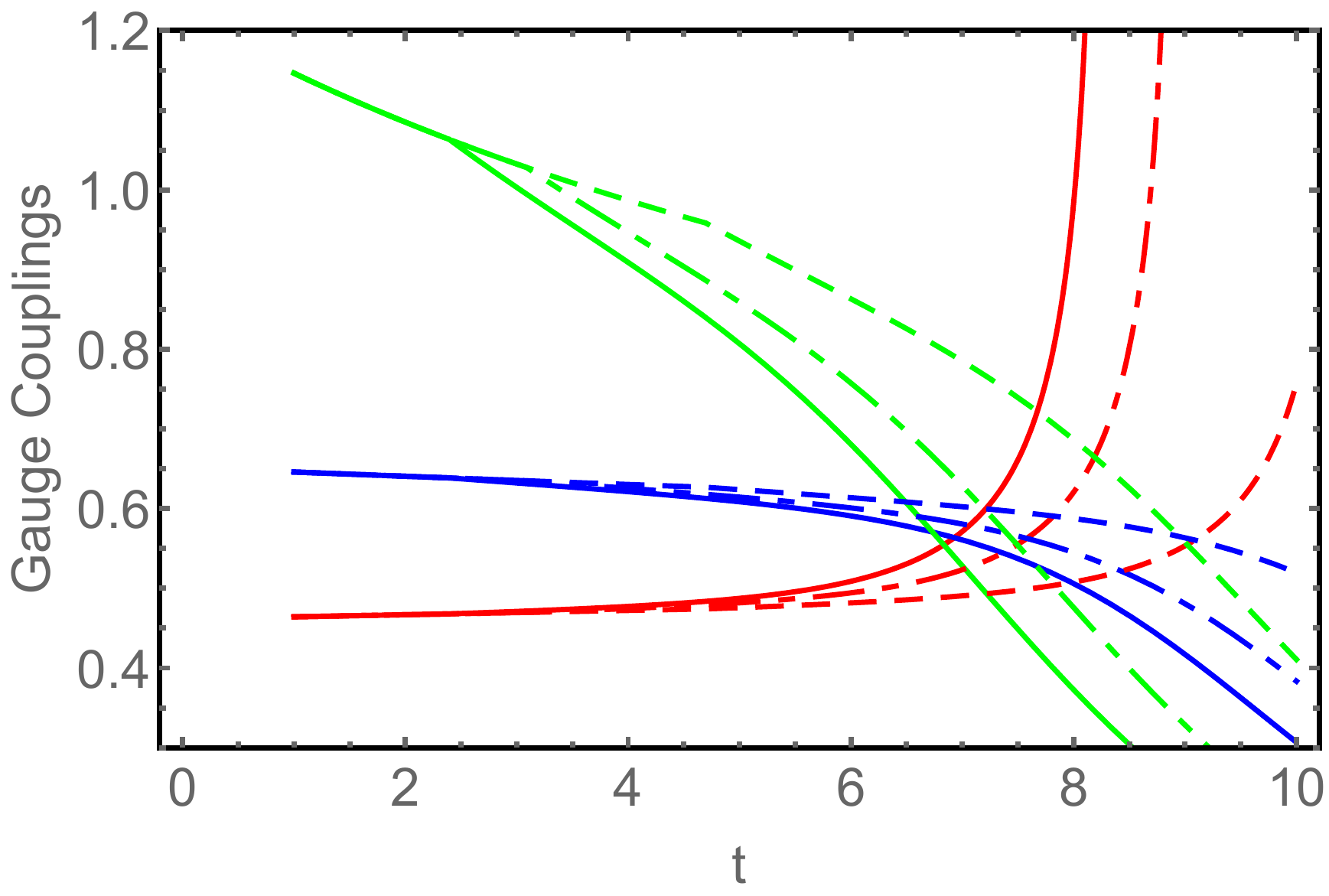}\hspace{1cm}
\includegraphics[width=6.5cm]{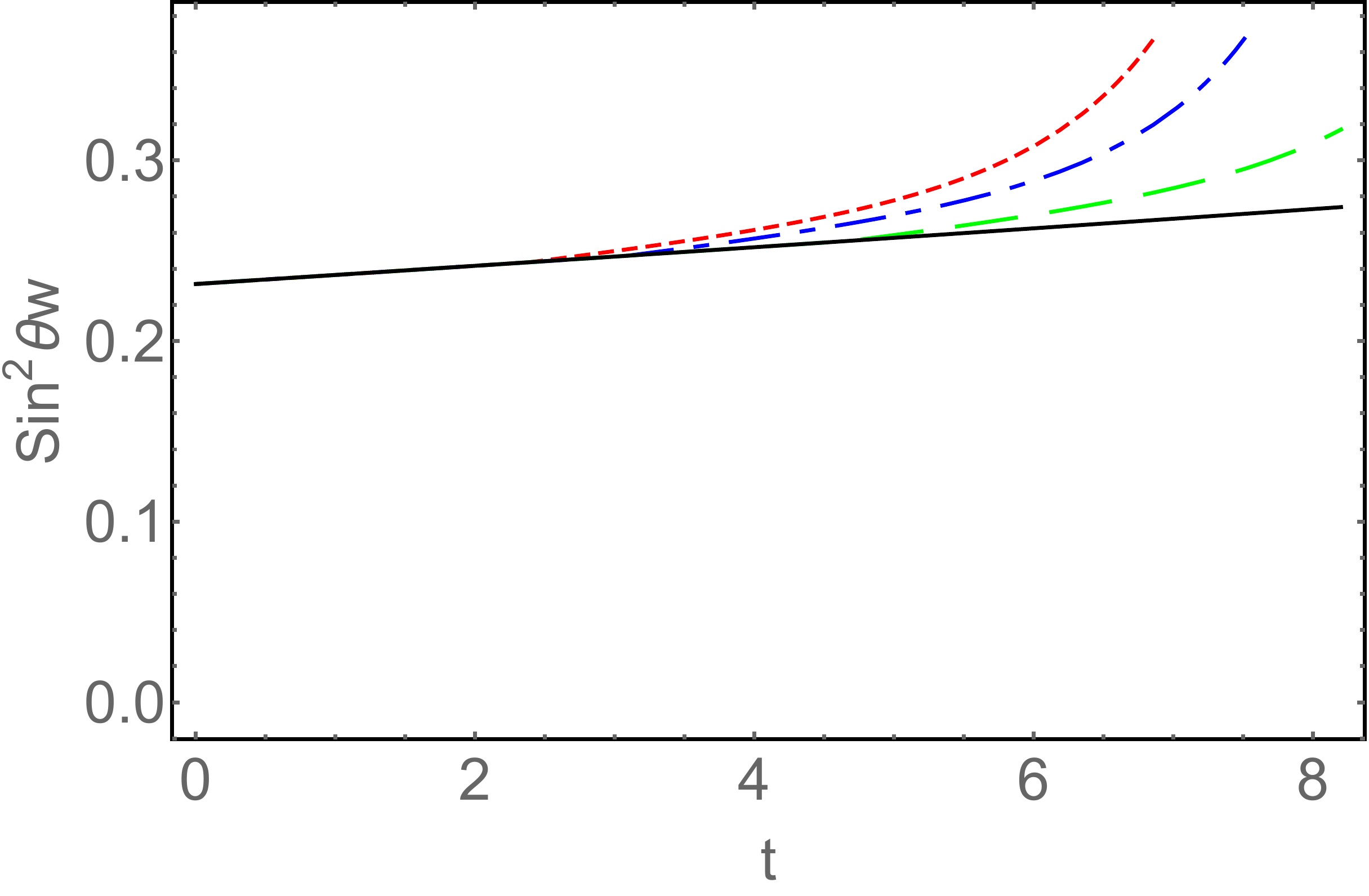}
\caption{ Left panel: Evolution of the gauge couplings $g_{1}$ (red), $g_{2}$ (blue) and $g_{3}$ (green), for three values of the $R^{-1}$ = 1 TeV (solid line), 2 TeV (dot-dashed line), 10 TeV (dashed line) as a function of $t$. Right panel: Evolution of the Weinberg angle  $\sin^{2} \theta_{W}$ with the bulk fermions, with the doublet located at position $y_{1}$ and two singlets located at position $y_{2}$, for $R^{-1} = 1$ TeV (red), $R^{-1} = 2$ TeV (blue) and $R^{-1} = 10$ TeV (green) as a function of $t$.}
\end{figure}
In Figure 1, left panel, we present the evolution of the gauge couplings for the one-loop $\beta$-functions, by assuming that the bulk fermion is the top quark. We see that the three gauge couplings unify at some value of $t$. For example, in the case of compactification scale $R^{-1}$= 1 TeV, there is approximate unfication at $t$=7. In the right panel we present the evolution of the Weinberg angle for the one loop $\beta$-functions, for different values of compactification scales, for the model discussed in the previous section. Recall that the Weinberg angle from group theory arguments for an $SU(3)$ gauge group is equal to $0.75$ (the ratio of couplings at high energy). In the right panel $\sin^{2}\theta_{W}\sim 0.69$ for the compactification scale $R^{-1}$= 1 TeV at scale parameter $t$ = 8. We can conclude that with this model the estimate of the Weinberg angle is closer to the group theoretically predicted Weinberg angle.

\section*{Acknowledgments}
This work is supported by the National Research Foundation (South Africa).

\section*{References}
\providecommand{\href}[2]{#2}\begingroup\raggedright\endgroup

\end{document}